\documentclass[prd,aps,preprintnumbers]{revtex4} 
\usepackage{amssymb}
\usepackage{graphicx}
\usepackage{axodraw}

\begin{document}

\title {\large Correlation between LFV and 
muon $(g-2)$ in MSSM}
 
\author{ Xiao-Jun Bi }
\affiliation{ Department of Physics, Tsinghua University, Beijing 100084,
People's Republic of China}
\email[Email: ]{bixj@mail.tsinghua.edu.cn}

\date{\today}

\begin{abstract}          

We give a simultaneous analysis of $g_\mu-2$ and lepton flavor violation
within MSSM. Working on the  interaction basis, we give direct relations
between $\delta a_\mu$ and Br$(\tau\to\mu\gamma)$ induced by the 
supersymmetric particles. We find the SUSY parameter space to satisfy
the $g_\mu-2$ constraint can be
considerably different from that given in case of no lepton flavor
mixing in the soft SUSY breaking sector.

\end{abstract}

\preprint{TUHEP-TH-02141}

\maketitle
 
Recently, the Brookhaven E821 Collaboration announced their new
experimental result on muon anomalous magnetic moment, $a_\mu=(g_\mu-2)/2$,
with improved statistics\cite{new821}.
The present discrepancy between the standard
model (SM) prediction and the measurement
is $a_\mu^{exp}-a_\mu^{SM} = 26(10)\times 10^{-10}$.

In this work we study the SUSY contributions to $a_\mu$ in the case
when considering the lepton flavor mixing in the soft SUSY breaking sector.
Besides the numerical studies of
lepton flavor mixing effects on $a_\mu$,
we give a thorough analysis of the correlation between
the SUSY contributions to $a_\mu$ and to lepton flavor violation (LFV).
We find that in this case
the SUSY parameter space may be quite different from that without
slepton mixing.

The effective Lagrangian related to $a_\mu$ and LFV
is as follow:
\begin{equation}
\label{eff}
{\cal L}_{eff}=e\frac{m_i}{2}\bar{l}_j\sigma_{\alpha\beta}F^{\alpha\beta}
(A_L^{ij}P_L+A_R^{ij}P_R)l_i\ ,
\end{equation}
where $i(j)$ denotes the initial (final) lepton flavor.
The $a_\mu$ is given by
\begin{equation}
 a_\mu=m_\mu^2(A_L^{22}+A_R^{22}),
\end{equation} 
while the branching ratio of
$\tau\to \mu\gamma$ is given by
\begin{equation}  
\text{Br}(\tau\to \mu\gamma)=\frac{\alpha_{em}}{4}m_\tau^5(|A^{23}_L|^2+|A^{23}_R|^2)/
\Gamma_\tau, 
\end{equation}          
with $\Gamma_\tau$ being the tau decay width.
We can see that the expressions for $\delta a_\mu$ and $\tau\to \mu\gamma$ are
closed related.

The SUSY contribution to the form factors $A_L$ and $A_R$ is given by
the photon-penguin diagrams via exchanging (i) chargino-sneutrino and (ii)
neutralino-slepton as shown in FIG. \ref{fig1}.
The analytic expressions for $\delta a_\mu$ can be found in Ref. \cite{bi}.
These expressions are given in the
mass eigenstates of the SUSY particles. It is suitable to do
numerical calculations on this basis. However, to analyze the
physical effects, it is more convenient to work on the {\em interaction
basis}, which is defined as the basis where the lepton mass
matrix and the gauge coupling vertices are diagonal. On this basis
the Feynman diagrams are more complicated than those in FIG. \ref{fig1}.

\begin{center}
\begin{figure}
\begin{picture}(500,250)(20,50)
\ArrowLine(20,200)(65,200)
\ArrowLine(65,200)(185,200)
\ArrowLine(185,200)(230,200)
\ArrowLine(270,200)(315,200)
\ArrowLine(315,200)(435,200)
\ArrowLine(435,200)(480,200)
\DashArrowArcn(125,200)(60,180,0){4}
\DashArrowArcn(375,200)(60,180,0){4}
\Photon(125,185)(180,130){4}{7}
\Text(150,135)[]{\huge $\gamma$}
\Text(170,250)[l]{\huge $\tilde{\nu}_\alpha$}
\Text(30,185)[]{\huge $\mu(\tau)$}
\Text(220,185)[]{\huge $\mu$}
\Text(125,215)[]{\Large $\chi^-_a$}
\Text(195,125)[]{\huge $q$}

\Photon(380,270)(450,305){4}{7}
\Text(440,285)[]{\huge $\gamma$}
\Text(328,250)[r]{\huge $\tilde{l}_\alpha$}
\Text(280,185)[]{\huge $\mu(\tau)$}
\Text(470,185)[]{\huge $\mu$}
\Text(375,185)[]{\Large $\chi^0_a$}
\end{picture}

\vspace*{-2.5cm}
\caption{\label{fig1} Feynman diagrams of the one-loop SUSY
contribution to $a_\mu$ (and the process $\tau \to \mu\gamma$)
via the exchange of a chargino (left) and via a neutralino (right).}
\end{figure}
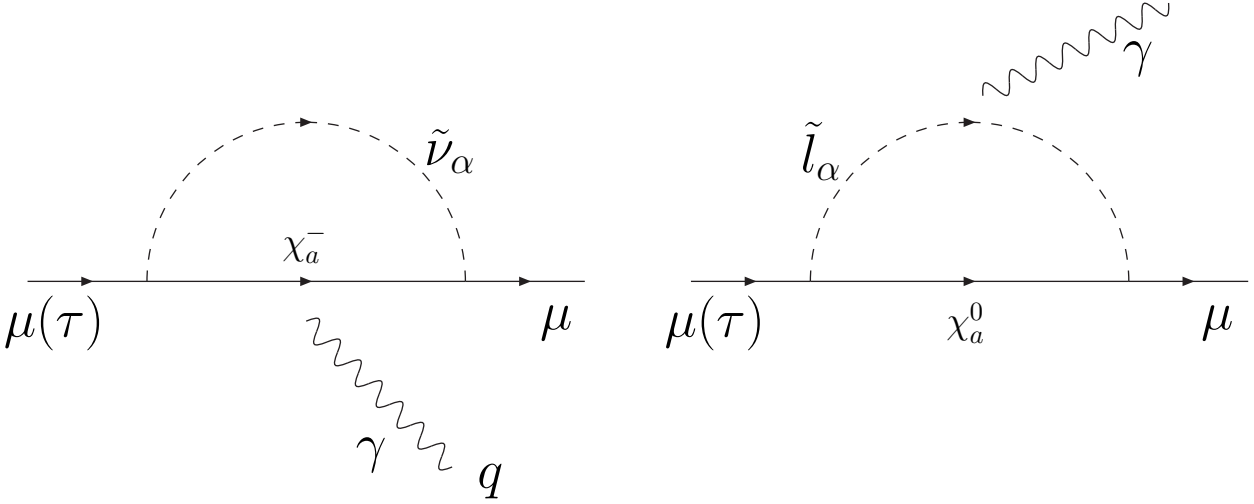
\end{center}

\vspace{-1cm}
On this basis the sneutrino and slepton mass matrices are
generally not diagonal. 
They can be written in a general form as  
\begin{equation}
M_{\tilde{\nu}}^2=Z_L m_L^2 Z_L^\dagger\ \ ,
\end{equation}
and
\begin{equation}
M_{\tilde{l}}^2=\left(  \begin{array}{cc}
Z_L m_L^2 Z_L^\dagger & -m_l(\mu\tan\beta+ A_l^*) \\
-m_l(\mu^*\tan\beta+ A_l) & 
Z_R m_R^2 Z_R^\dagger  
\end{array} \right)\ \ ,
\end{equation}
where
$Z_L$ and $Z_R$ are the left- and right-handed mixing matrices.
In this work we consider the mixing between the second
and the third generations. Thus, $Z_L$ and $m_L^2$ are given by(similar for
right-handed parameters)
\begin{equation}
Z_L=\left( \begin{array}{cc}  \cos\theta_L & \sin\theta_L \\
  -\sin\theta_L &\cos\theta_L \end{array}\right)\ \ \text{and}\ \ \ 
m_L^2 = \left( \begin{array}{cc} m_2^2 & \\ & m_3^2 \end{array}\right)\ . 
\end{equation}

Since on this basis all the interaction vertices are diagonal,
the LFV effects are transferred to the propagators of sleptons and
sneutrinos. The propagator of a scalar is $\frac{i}{p^2-M^2}$.
In our case with $p^2 \ll M^2$, the slepton and sneutrino 
propagators can be approximately
given by the inverse of the corresponding mass matrices. It is
easy to get
\begin{equation} 
(M_{\tilde{\nu}}^2)^{-1} =\left( \begin{array}{cc}
\frac{c_L^2}{m_2^2}+\frac{s_L^2}{m_3^2} & c_Ls_L\frac{m_2^2-m_3^2}{m_2^2m_3^2} \\
c_Ls_L\frac{m_2^2-m_3^2}{m_2^2m_3^2} & \frac{s_L^2}{m_2^2}+\frac{c_L^2}{m_3^2}
\end{array}  \right)\ \ .
\end{equation}
$\delta a_\mu$ is related to the propagator of $\tilde{\nu}_\mu-\tilde{\nu}_\mu$,
which is noted as
$P(\tilde{\nu}_\mu-\tilde{\nu}_\mu)=\frac{c_L^2}{m_2^2}+\frac{s_L^2}{m_3^2}$,
while $\tau\to\mu\gamma$
is related to the propagator of $\tilde{\nu}_\tau-\tilde{\nu}_\mu$, 
which is 
$P(\tilde{\nu}_\tau-\tilde{\nu}_\mu)=
\frac{1}{2}\sin2\theta_L\frac{m_2^2-m_3^2}{m_2^2m_3^2}$.

We consider the following two limit cases:
\begin{equation}
P(\tilde{\nu}_\mu-\tilde{\nu}_\mu)\rightarrow \left\{
\begin{array}{c} m_2^2\approx m_3^2 \approx m^2,\ \ \frac{1}{m^2} \\
m_2^2 \gg m_3^2,\ \ \frac{s_L^2}{m_3^2}
\end{array} \right.\ \ ,
\end{equation}
while
\begin{equation}
P(\tilde{\nu}_\tau-\tilde{\nu}_\mu)\rightarrow  \left\{
\begin{array}{l} m_2^2\approx m_3^2,\ \ \sim 0 \\
m_2^2 \gg m_3^2,\ \ \frac{1}{2}\sin2\theta_L\frac{1}{m_3^2}
\end{array} \right.\ \ .
\end{equation}
In the first case with $m_2^2\approx m_3^2$, we can see that $\delta a_\mu$ 
does not depend on the mixing angle $\theta_L$ while Br$(\tau\to\mu
\gamma)$ tends to zero. Thus models with gravity or
gauge mediated supersymmetry breaking may predict that $\delta a_\mu$ has
nothing to do with the mixing angle $\theta_L$ while Br$(\tau\to\mu\gamma)$ 
should be very small\cite{bi}. Thus the first case is actually the same as the case
of no lepton flavor mixing in the soft sector. The second case leads us
to the effective SUSY scenario\cite{effe}, where the first two generations' sfermions
are as heavy as about $20 TeV$. In this work we mainly consider the latter case.
In this case the two quantities are closely correlated.

The propagator of slepton are approximately given by
\begin{equation}
(M_{\tilde{l}}^2)^{-1} \approx \left( \begin{array}{cc} A&C\\ C&B\end{array}\right)\ ,
\end{equation}
with
\begin{equation}
A\approx (M_{\tilde{\nu}}^2)^{-1},\ \ B\approx A(\theta_L\to\theta_R)
\end{equation}
and
\begin{equation}
C\approx m_\tau\mu\tan\beta\frac{m_2^2-m_3^2}{m_2^2m_3^2}\cdot 
\left[ \begin{array}{ll} \frac{1}{4}\sin2\theta_L\sin2\theta_R
\frac{m_2^2-m_3^2}{m_2^2m_3^2} & \frac{1}{2}\sin2\theta_L
\left( \frac{s_R^2}{m_2^2}+\frac{c_R^2}{m_3^2} \right) \\
\frac{1}{2}\sin2\theta_R \left( \frac{s_L^2}{m_2^2}+\frac{c_L^2}{m_3^2}\right) &
\left(\frac{s_L^2}{m_2^2}+\frac{c_L^2}{m_3^2} \right)
\left( \frac{s_R^2}{m_2^2}+\frac{c_R^2}{m_3^2}\right)
 \end{array} \right]\ \ .
\end{equation}
In matrix $C$ we have omitted the terms proportional to $m_\mu$.
From the above expressions we know the propagator of
$\tilde{\mu}_L-\tilde{\mu}_L$ is the same as that of
$\tilde{\nu}_\mu-\tilde{\nu}_\mu$.
The most interesting result is the propagator of 
$\tilde{\mu}_L-\tilde{\mu}_R$, given by
\begin{equation}
P(\tilde{\mu}_L-\tilde{\mu}_R)=
\frac{1}{4}m_\tau\mu\tan\beta\sin2\theta_L\sin2\theta_R
\left(\frac{m_2^2-m_3^2}{m_2^2m_3^2}\right)^2\ \ ,
\end{equation}
which has an $m_\tau$ enhancement. This term may dominate over
others if both the left- and right-handed mixing
angles are large. \\

{\bf\large $\delta a_\mu$ with only left-handed mixing }\\

\begin{center}
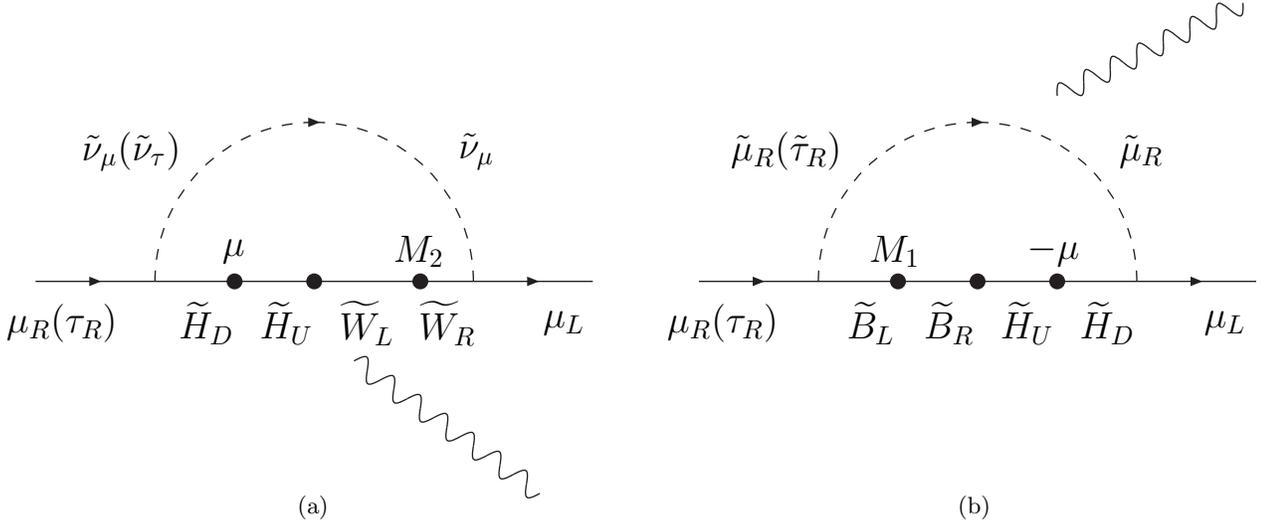
\begin{figure}
\begin{picture}(500,250)(20,50)
\ArrowLine(20,200)(65,200)
\ArrowLine(65,200)(185,200)
\ArrowLine(185,200)(230,200)
\DashArrowArcn(125,200)(60,180,0){4}
\Photon(140,170)(210,120){4}{7}
\ArrowLine(270,200)(315,200)
\ArrowLine(315,200)(435,200)
\ArrowLine(435,200)(480,200)
\DashArrowArcn(375,200)(60,180,0){4}
\Photon(405,270)(475,305){4}{7}
\Text(75,250)[r]{\Large $\tilde{\nu}_\mu(\tilde{\nu}_\tau)$}
\Text(180,250)[l]{\Large $\tilde{\nu}_\mu$}
\Text(30,185)[]{\Large $\mu_R(\tau_R)$}
\Text(220,185)[]{\Large $\mu_L$}
\Text(95,185)[r]{\Large $\widetilde{H}_D$}
\Text(125,185)[r]{\Large $\widetilde{H}_U$}
\Text(135,185)[l]{\Large $\widetilde{W}_L$}         
\Text(165,185)[l]{\Large $\widetilde{W}_R$}
\Text(95,212)[]{\Large $\mu$}
\Text(165,212)[]{\Large $M_2$}
\Vertex(95,200){3}
\Vertex(125,200){3}
\Vertex(165,200){3}
\Text(125,115)[]{(a)}

\Text(325,250)[r]{\Large $\tilde{\mu}_R(\tilde{\tau}_R)$}
\Text(430,250)[l]{\Large $\tilde{\mu}_R$}
\Text(280,185)[]{\Large $\mu_R(\tau_R)$}
\Text(470,185)[]{\Large $\mu_L$}
\Text(345,185)[r]{\Large $\widetilde{B}_L$}
\Text(375,185)[r]{\Large $\widetilde{B}_R$}
\Text(385,185)[l]{\Large $\widetilde{H}_U$}
\Text(415,185)[l]{\Large $\widetilde{H}_D$}
\Text(345,212)[]{\Large $M_1$}
\Text(405,212)[]{\Large $-\mu$}
\Vertex(345,200){3}
\Vertex(375,200){3}
\Vertex(405,200){3}
\Text(375,115)[]{(b)}
\end{picture}

\vspace*{-2.5cm}
\caption{\label{fig2} Feynman diagram which gives the dominant
contribution to $\delta a_\mu$ (and to the process $\tau \to\mu\gamma$)
in case of (a) only left-handed slepton mixing and (b) only right-handed mixing. 
The black dots in the fermion
line are mass insertions.}
\end{figure}
\end{center}

\vspace{-1cm}
When there is only left-handed mixing, the most important
contribution to $\delta a_\mu$ and Br$(\tau\to\mu\gamma)$ 
comes from the diagram in FIG. \ref{fig2}(a), given on the
interaction basis. From this diagram
we can directly read that
\begin{equation}
A_R^{23}(c)=\frac{1}{2}\delta a_\mu^{(c)}/m_\mu^2\frac{Z_L^{33}}{Z_L^{23}}\ \ \ .
\end{equation}
Then we have, assuming $\theta_L=\pi/4$, that
\begin{eqnarray}
Br(\tau\to\mu\gamma)&\approx & \frac{\alpha_{em}}{4}m_\tau^5|A_R^{23}(c)|^2/\Gamma_\tau \nonumber \\
&\approx & 2.9\times 10^{13} |\delta a_\mu|^2\ .
\end{eqnarray}
From the present upper limit of $Br(\tau\to\mu\gamma) < 10^{-6}$,
we get that 
\begin{equation}
\delta a_\mu < 1.9 \times 10^{-10},\ \ \text {in case of}\ \ \theta_R=0\ .
\end{equation}

From this diagram we also have
the conclusion that
\begin{equation}
\mu M_2 > 0, \ \ \text {in case of}\ \ \theta_R=0\ 
\end{equation}
to give positive contribution to  $\delta a_\mu$.
The same diagram gives the dominant contribution to $\delta a_\mu$
in the case of no lepton flavor mixing. Thus the same conclusion
of the sign of $\mu$ is given in that case.\\

{\bf \large $\delta a_\mu$ with only right-handed mixing } \\

In case of only right-handed mixing, the chargino-sneutrino diagram
gives no contribution to $\delta a_\mu$. The most important
contribution to $\delta a_\mu$ and Br$(\tau\to\mu\gamma)$
comes from the diagram in FIG. \ref{fig2}(b), given on the
interaction basis. 
If we ignore the mixing between the left- and right-handed sleptons,
$Z_R$ is approximately the slepton mixing matrix.
From FIG. \ref{fig2}(b) we then have 
\begin{equation}
\label{rmix}
A_R^{23}(n)\approx \frac{1}{2}\frac{\delta a_\mu^{(n)}}{m_\mu^2}
\left(\frac{m_\mu}{m_\tau}\right)
\frac{Z_R^{33}}{Z_R^{23}}\ \ \ .
\end{equation}

Then we have, assuming $\theta_R=\pi/4$, that
\begin{eqnarray}
Br(\tau\to\mu\gamma)&\approx & \frac{\alpha_{em}}{4}m_\tau^5|A_R^{23}(n)|^2/\Gamma_\tau \nonumber \\
&\approx & 1.\times 10^{11} |\delta a_\mu|^2\ .
\end{eqnarray}
From the present upper limit of $Br(\tau\to\mu\gamma) < 10^{-6}$,
we get that
\begin{equation}
\delta a_\mu < 32 \times 10^{-10},\ \ \text {in case of}\ \ \theta_L=0\ .
\end{equation}

This upper bound is much larger than that in case of only left-handed
mixing. It is obvious that the factor $\frac{m_\mu}{m_\tau}$ in 
Eq. (\ref{rmix}), which greatly suppresses Br($\tau\to\mu\gamma$), 
helps to increase the bound.
This factor comes  from the Yukawa vertex on the right, where the Higgsino 
component $\widetilde{H}_D$
 has to be associated
with the muon line since there is only right-handed mixing in the slepton sector.
However, in  FIG. \ref{fig2}(a), where the charged
Higgsino component $\widetilde{H}_D$ is associated with
the tau line, no such factor helps to suppress Br($\tau\to\mu\gamma$).

Another interesting point is that the mass insertion for the
neutral component of  $\widetilde{H}_U \widetilde{H}_D$ is
$-\mu$, while it is $\mu$ for the same term of the charged component.
Thus we have 
\begin{equation}
\mu M_1 < 0, \ \ \text {in case of}\ \ \theta_L=0\
\end{equation}
to give positive contribution to  $\delta a_\mu$.
This means that if we set $M_1$ and $M_2$ have same sign,
which is well motivated theoretically, $\mu$ should be
negative in this case. 

\begin{figure}
\includegraphics[scale=0.4]{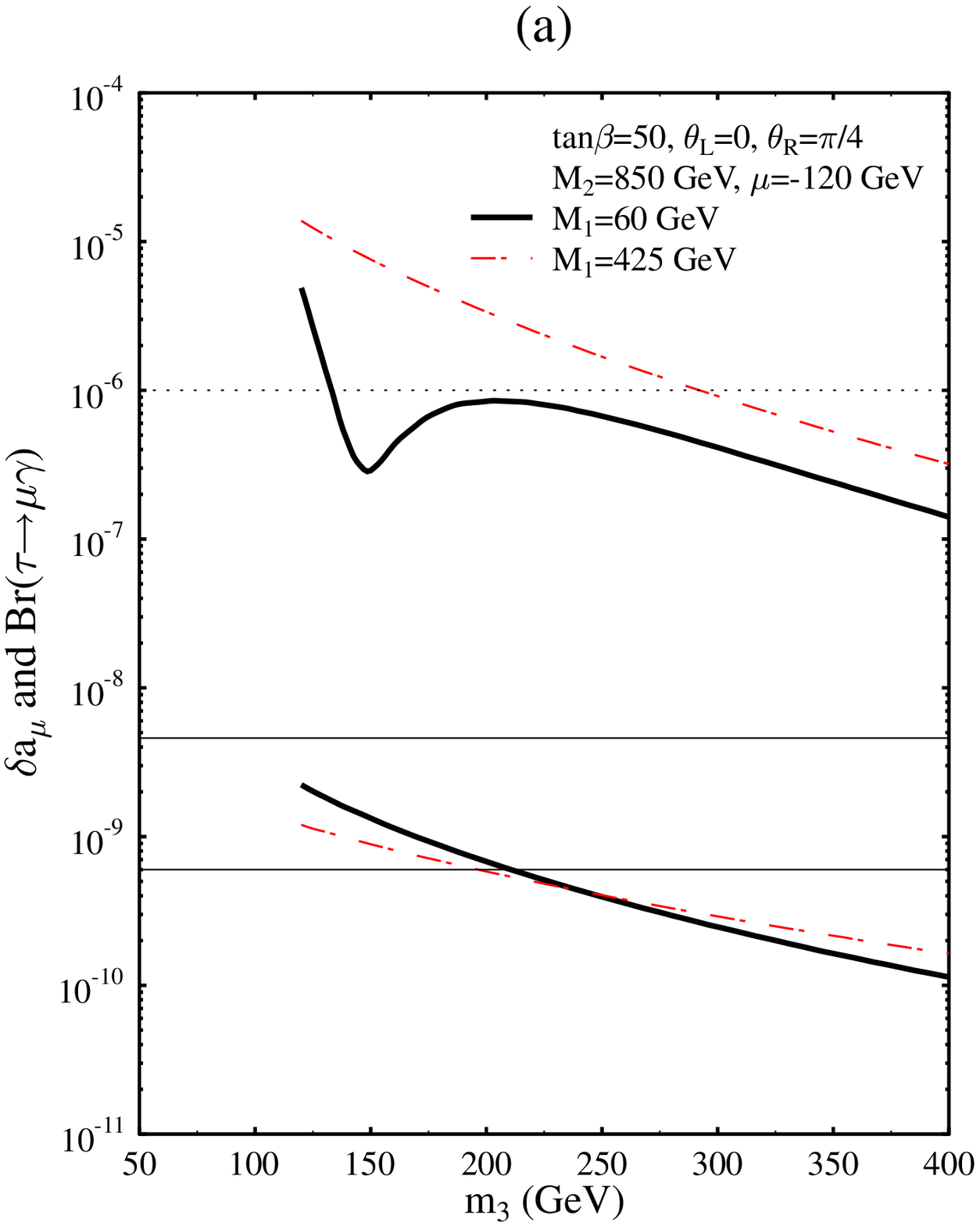}
\includegraphics[scale=0.4]{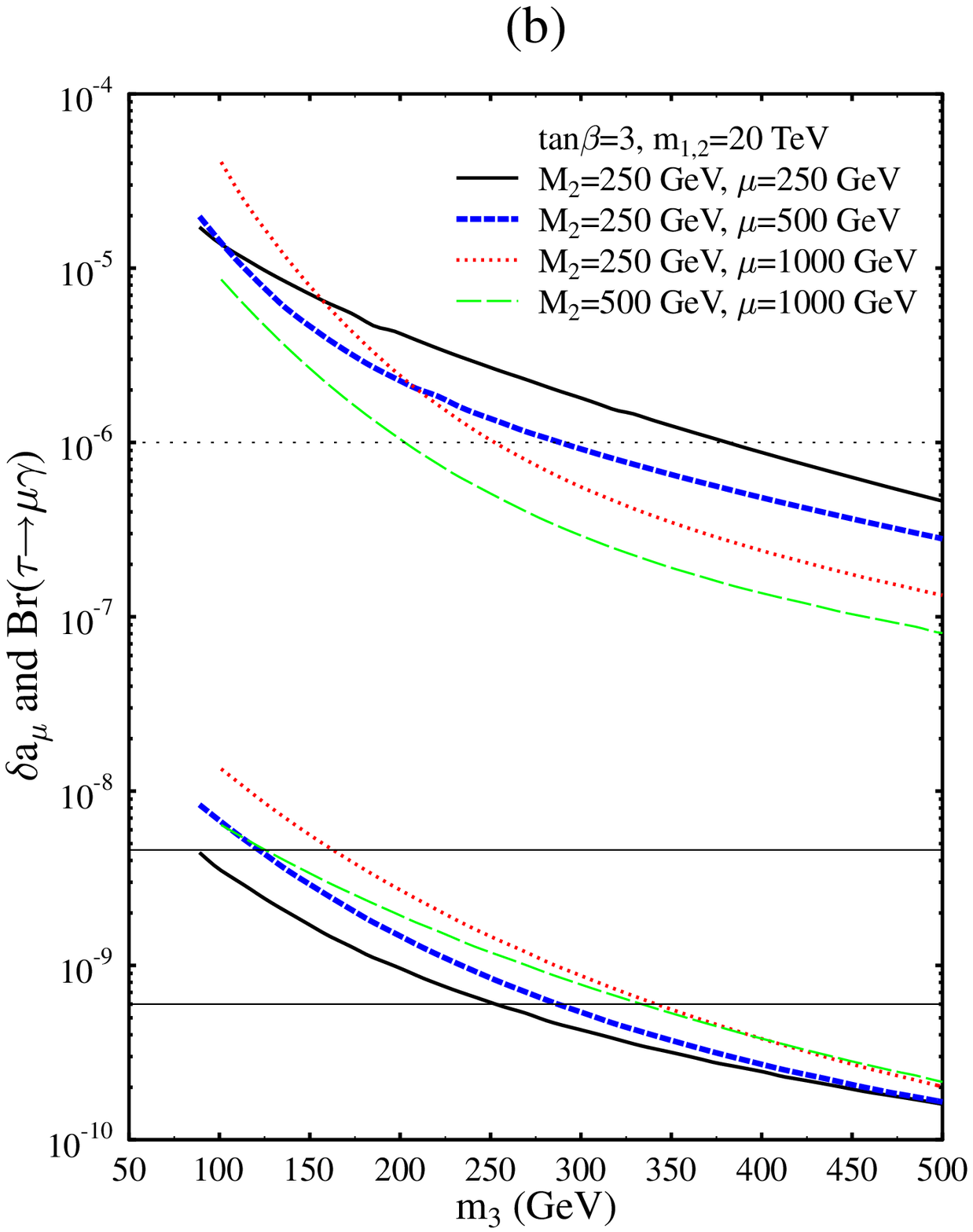}
\caption{\label{fig3} 
$\delta a_\mu$ and Br($\tau\to\mu\gamma$) as functions of $m_3$ in
case of (a) $\theta_L=0$, $\theta_R=\pi/4$ and (b) $\theta_L=\pi/4$, 
$\theta_R=\pi/4$. 
The horizontal lines represent the E821 $\pm 2\sigma$ 
bounds (solid) and the upper limit of Br($\tau\to\mu\gamma$) (dotted). }
\end{figure}

Since we ignore the left-right mixing  between the sleptons,
the naive bound we get should be examined numerically.
In FIG. \ref{fig3}(a) we show the numerical results in this case.
If we adopt the GUT motivated relation 
$M_1=\frac{5}{3}\frac{\alpha_1}{\alpha_2}M_2\approx 0.5M_2$
we have $\delta a_\mu < 3.\times 10^{-3}$ to satisfy the 
Br($\tau\to\mu\gamma$) bound.
However, if we relax the above relation and fix $M_1=60 GeV$, $\delta a_\mu$
can be as large as $\sim 17\times 10^{-10}$ without violating
the bound of Br($\tau\to\mu\gamma$).\\

{\bf\large $\delta a_\mu$ with both left- and right-handed mixing } \\

\begin{center}
\begin{figure}
\begin{picture}(500,250)(20,50)
\ArrowLine(145,200)(190,200)
\ArrowLine(190,200)(310,200)
\ArrowLine(310,200)(355,200)
\DashArrowArcn(250,200)(60,180,0){4}
\Photon(280,270)(350,305){4}{7}
\Text(155,185)[]{\Large $\mu_R$}
\Text(345,185)[]{\Large $\mu_L$}
\Text(220,185)[]{\Large $\widetilde{B}_L$}
\Text(290,185)[]{\Large $\widetilde{B}_R$}
\Text(180,230)[]{\Large $\tilde{\mu}_R$}
\Text(310,230)[l]{\Large $\tilde{\mu}_L$}
\Vertex(250,200){3}

\end{picture}

\vspace*{-4.cm}
\caption{\label{fig4}
Feynman diagram which gives the dominant
contribution to $\delta a_\mu$
in the case that both the left- and right-handed slepton mixing
are large. }
\end{figure}
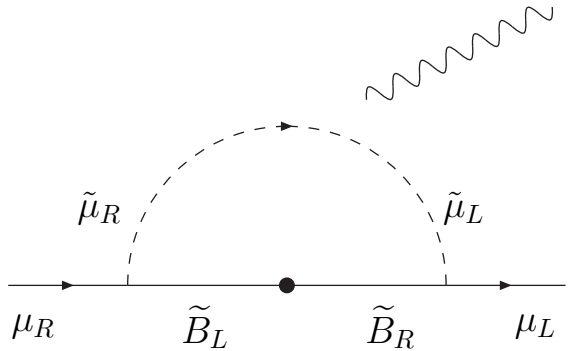
\end{center}

\vspace{-1cm}
In this case we have derived that there is an $m_\tau$ enhancement
in the propagator $\tilde{\mu}_L-\tilde{\mu}_R$.
The enhancement leads to that the diagram in FIG. \ref{fig4}
may give dominant contribution to $\delta a_\mu$ if both $\theta_L$ and
 $\theta_R$ are large. However, it seems that there is no obvious
term which give dominant contribution to Br($\tau\to\mu\gamma$).
If $m_3$ is small, FIG. \ref{fig4} with $\mu_R$ replaced by $\tau_R$ may
dominates other terms. In this case we get a similar
limit as that given in the case with only right-handed mixing,
\begin{equation}
\delta a_\mu < 32\times 10^{-10} \ \ \ \text{in case of no}\ \ \theta=0\ \ .
\end{equation}
However, this bound is very loose because in large parameter space
the contribution to Br($\tau\to\mu\gamma$) by exchanging $\chi^-$
is more important than that by exchanging $\chi^0$. We have to
study this case numerically.

FIG. \ref{fig3}(b) displays $\delta a_\mu$ and Br($\tau\to\mu\gamma$)
with $\theta_L=\theta_R=\pi/4$. If $M_2$ and $\mu$ are both large,
there is a large region which can accommodate 
$\delta a_\mu$ and Br($\tau\to\mu\gamma$) simultaneously.
As $\mu$ becomes large,  Br($\tau\to\mu\gamma$) decreases
while $\delta a_\mu$ increases. This is understood that
large $\mu$ enhances the propagator of $P(\tilde{\mu}_L-\tilde{\mu}_R)$ 
and leads to large chargino mass, which decreases Br($\tau\to\mu\gamma$).

In summary, when both the left- and right-handed slepton
mixing is large, SUSY can enhance $\delta a_\mu$ to within
the E821 $\pm 2\sigma$ bounds in a large parameter space
through the slepton mixing between the second and the third
generations. In this case small $\tan\beta$ is slightly favored.
Higgsino mass parameter $\mu$ can be either positive or negative
depending on the relative sign between $\theta_L$ and $\theta_R$.
$\delta a_\mu$ can reach up to $\sim 20\times 10^{-10}$ even
keeping the relation $M_1\approx 0.5 M_2$, which means bino
is not necessarily kept very light.

\begin{acknowledgments}
 
This work is supported by the National Natural Science Foundation
of China under the grand No. 10105004.
 
\end{acknowledgments}

\end{document}